\documentclass[galaxies,article,accept,moreauthors,pdftex]{mdpi}
\usepackage{comment} 

\firstpage{1} 
\pubvolume{xx}
\issuenum{1}
\articlenumber{5}
\pubyear{2020}
\copyrightyear{2020}
\history{Received: 27 February 2020; Accepted: 28 April 2020; Published: date}
\updates{yes} % if~there is~an update available, un-comment this line

\Title{The Interaction of Type Ia Supernovae with Planetary Nebulae: the~Case of Kepler's Supernova~Remnant}
\Author{{A. Chiotellis} *, P. Boumis  and~Z. T. Spetsieri }
\AuthorNames{A. Chiotellis, P. Boumis and~Z. T. Spetsieri }

\address [1] {%
Institute for Astronomy, Astrophysics, Space Applications and~Remote Sensing, National Observatory of Athens, 15236 Penteli, Athens, Greece; ptb@noa.gr (P.B.); zspetsieri@noa.gr (Z.T.S.)}
\corres{Correspondence: a.chiotellis@noa.gr}

\abstract{One of the~key methods for determining the~unknown nature of Type Ia supernovae (SNe~Ia) is~the~search for traces of interaction between the~SN ejecta and~the~circumstellar structures at the~resulting supernova remnants (SNRs Ia). So far, the~observables that we~receive from well-studied SNRs Ia cannot be~explained self-consistently by any model presented in~the~literature. In~this study, we~suggest that the~circumstellar medium (CSM) being observed to~surround several SNRs~Ia was mainly shaped by planetary nebulae (PNe) that originated from one or~both progenitor stars. Performing two-dimensional hydrodynamic simulations, we~show that the~ambient medium shaped by PNe can account for several properties of the~CSM that have been found to~surround SNe~Ia and~their remnants. Finally, we~model Kepler's SNR considering that the~SN explosion occurred inside a~bipolar PN. Our simulations show good agreement with the~observed morphological and~kinematic properties of Kepler's SNR. In particular, our model reproduces the~current expansion parameter of Kepler's SNR, the~partial interaction of the~remnant with a~dense CSM at its northern region and~finally the~existence of two opposite protrusions (`ears') at the~equatorial plane of the~SNR.}

\keyword{type Ia supernovae; supernova remnants; planetary nebulae; SN 1604, Kepler's SNR}

\begin{document}
%%%%%%%%%%%%%%%%%%%%%%%%%%%%%%%%%%%%%%%%%%

%%%%%%%%%%%%%%%%%%%%%%%%%%%%%%%%%%%%%%%%%%
\section{Introduction}

Thermonuclear or~Type Ia supernovae (SNe~Ia) is~a~class of supernovae that results from the~thermonuclear combustion of a~carbon-oxygen white dwarf (WD), which is~destabilized through the~interaction with a~companion star. This interaction could appear either via mass transfer from a~non-degenerate donor star (single degenerate scenario, SD) or~in~a~merger event with a~degenerate companion (double degenerate scenario, DD). However, both scenarios have severe weaknesses and~cannot account for all the~observed properties of SNe Ia~\cite{Livio18}. As~a~result, despite decades of research the~nature of the~donor star, the~mass transfer process and~the~explosion mechanism itself remain highly uncertain. Given that SNe~Ia have fundamental consequences in~a~wide range of astrophysical issues (e.g., low mass binary evolution, chemical enrichment of galaxies, cosmology), the~identification of their unknown nature is~considered to be one of the~most crucial quests of stellar~astrophysics. 

\textls[-25]{A promising method that attempts to~clarify the~unknown origin of these cosmic explosions is~the~study of the~SNe Ia aftermath, the~Supernova Remnants (hereafter SNRs Ia).} There~is~a~consensus that the~morphology, kinematics and~emission properties of SNRs reflect the~interaction of the~SN blast wave with inhomogeneous circumstellar medium (CSM) shaped by the~mass outflows of the~progenitor~system. \textls[-25]{Thus, deciphering the~observational data of nearby SNRs Ia, through detailed modeling, we~get crucial insights into the~structure and~the~properties of the~CSM that was surrounding the~explosion center and~extract valuable information about the~nature and~evolution of the~progenitor~system. }

Indeed, several well-studied SNRs Ia exhibit peculiar properties that are~only explained by assuming that the~SNR is~interacting with circumstellar structures formed by its parent stellar system. The~most characteristic case is~the~remnant of the~historical SN Ia observed by Kepler (SN~1604), which reveals profound evidence of interaction with dense CSM in~its northern and~central regions~\cite{Burkey13, Blair91}. A~number of models have been computed aiming to~reproduce the~observed properties of Kepler's SNR and~the~general conclusion is~that the~CSM around Kepler is~largely made of material expelled by a~wind-losing donor star in~the~Asymptotic Giant Branch (AGB), member of the~parent stellar system~\mbox{\cite{Burkey13,Patnaude12,Chiotellis12}}. However, no evidence of the~presence of a~survived AGB star has been found in~the~center of Kepler's SNR~\cite{Kerzendorf14} and~thus, a~different progenitor model of Kepler's SNR is~required. Discrepancies between theoretical predictions and~observations are~also found in~the~SNRs Ia of Tycho (SN 1572) and~RCW86, which seem to~expand in~an extended low density cavity~\cite{Zhou16,Broersen14}. The~formation of this cavity is~attributed to~the~mass outflows that emanate from the~surface of accreting WDs known as `accretion winds'~\cite{Hachisu96}. However, the~existence of accretion winds has been questioned in~the~literature~(\cite{Livio18}, and~references therein), while for the~case of Tycho's SNR such a~scenario does not agree with its observed X-ray spectra~\cite{Badenes07} and~a~steadily nuclear burning WD progenitor has been recently ruled out~\cite{Woods17}. In~conclusion, the~evidence and~constraints  posed by the~observation of several well-studied SNRs Ia rule out essentially any model suggested in~the~literature. 

In this work, we~propose an innovative scenario according to~which---at least a~fraction of---SNe Ia occur in~and~subsequently interact with Planetary Nebulae (PNe) formed by the~progenitor system (see also~\cite{Tsebrenko11}). We~demonstrate, using hydrodynamical simulations, that this model can account for diverse CSM properties observed to~surround several SNRs Ia. Finally, we~model the~SNR of Kepler within the~framework of the~studied model and~we show that the~interaction of the~SN ejecta with a~surrounding bipolar PN reproduces the~overall morphological and~kinematic properties of the~remnant.

%%%%%%%%%%%%%%%%%%%%%%%%%%%%%%%%%%%%%%%%%%

\section{Hydrodynamic Modeling of the~CSM Shaped by Planetary~Nebulae}\label{sec:PN_CSM}

PNe are~formed in~the~final stages of low mass stars ($M \le 8~ M_{\odot} $) which end their lives as WDs. That means that both SNe Ia and~PNe share a~common evolutionary path. In~addition, observational surveys reveal that most PNe central stars are~low mass binaries involving one or~two WDs~\cite{DeMarco13,Miszalski2009} i.e.,~as expected to~be the~progenitors of SNe Ia. Thus, it~is~most likely that PNe in~the~past contributed to~modify the~properties of the~CSM around the~progenitor~systems.

 In~order to~investigate the~ability of PNe to~reproduce the~CSM properties observed around well-studied SNRs Ia, we~employ the~hydrodynamic code AMRVAC~\cite{Keppens03}, and~we perform a~series of simulations modeling the~SNe Ia ambient medium shaped by a~PN. The~extracted results are~mainly determined by two parameters: a) the~characteristics of the~host PN and~b) the~time delay between the~PN formation (i.e., the~moment where the~newly born WD starts to ionize the~surrounding CSM) and~the~subsequent SN Ia explosion ($\tau_{delay}$). Given that PNe reveal a~vast diversity in~terms of size, shapes, kinematics etc.~\cite{Kwok18} while the~time delay between the~WD formation and~the~final SN Ia explosion can be~from almost zero up to~several Gyr \citep{Ruiter2009}, the~parameter space involved in~our modeling is~immense. Here, as~a~first attempt, we~consider the~case of a~{\it bipolar PN} as this morphology is~one the~most common PN morphologies. Regarding the~delay time between the~PN formation and~the~subsequent SN Ia explosion, we~present three cases each of which represents a~specific state of the~CSM around the~explosion center: (a) the~SN Ia occurs almost simultaneously with the~birth of the~WD i.e.,~the $\tau_{delay}$ is~negligible compared to~the~dynamical timescale of the~surrounding PN evolution ($\tau_{delay}$ = 0 Myr ), (b) the~WD explodes $\tau_{delay}$ = 2~Myr after its formation and~(c) the~time delay between the~two phenomena is~$\tau_{delay}$ = 8~Myr.

\begin{figure}[H]
\centering
\includegraphics[width=14 cm]{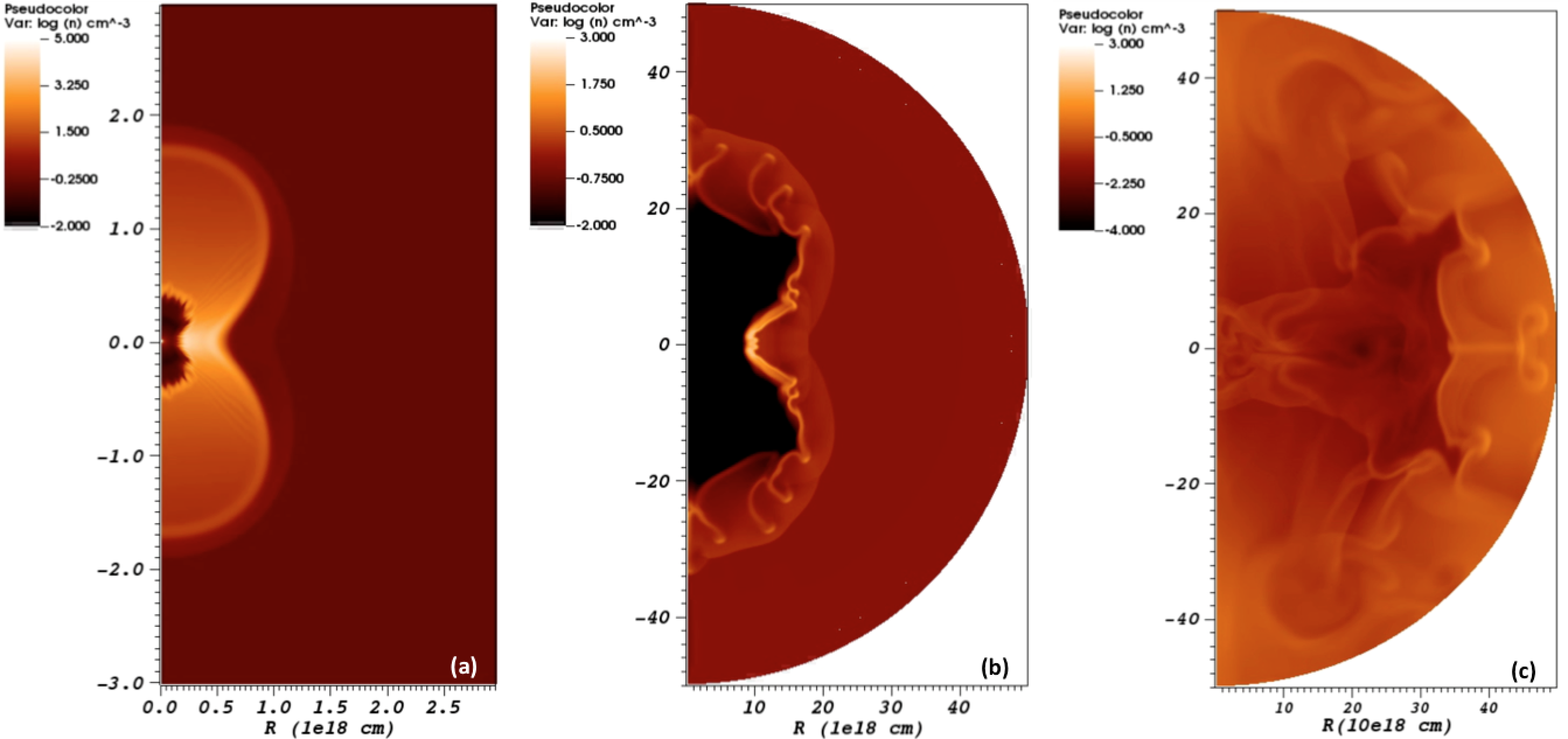}
\caption{The 2D density contours of the~CSM shaped by a~bipolar PN for the~cases where the~SN Ia explosion occurred right after the~cessation of the~fast wind (\textbf{a}), 2 Myr later (\textbf{b}) and~8 Myr later (\textbf{c}).}
\label{fig:CSM_PN}
\end{figure}

We perform our modeling on a~2D spherical grid and~assume symmetry in~the~third dimension. {Adopting the~interactive-stellar-wind model~\cite{Paczynski71,Kwok78,Balick1987} according to~which the~PNe shell are~shaped by the~interaction of the~fast wind from the~central star and~the~remnant of the~slow AGB wind, we~run our simulations in~two steps: (a)  we~first simulate the~formation of the~AGB wind bubble by imposing a~continuous inflow in~the~inner boundary of our grid in~the~form of a~slow stellar wind with $\dot{M} = 2 \times 10^{-5}~ \rm M_{\odot}~yr^{-1}$ and~$u_w= 5$~km~s$^{-1}$,  (b) after 0.5 Myr we~change the~inflow properties at the~inner boundary of the~grid and~we insert a~fast tenuous wind of $\dot{M} = 10^{-8}~ \rm M_{\odot}~yr^{-1}$ and~$u_w= 4 \times 10^3$~km~s$^{-1}$ in~the~computational domain.} The~fast wind starts to~sweep up the~previous CSM and~forms an inner cavity surrounded by a~shell (i.e.,~the~PN). The~wind's bipolarity is~described following the~trigonometric functions: $\dot{M}(\theta)= \dot{M}(0) \left[1 - \beta \mid \sin \theta \ \mid ^{k} \right]^{-1} $ and~$u_{w}(\theta) = u(0) \left[1 - \alpha \mid \sin \theta \ \mid ^{k} \right]$ where $\dot{M}(\theta)$ and~$u_{w}(\theta)$ is~the~wind mass loss rate and~terminal velocity at the~polar angle $\theta$, respectively (see {\cite{Huarte-Espinosa2012,Garcia-Segura1999,Zhang1998} for relevant  approaches on hydro-modeling of axisymmetric PNe).} The~$\alpha, \beta$ and~$k$ are~constants which determine the~polar distribution of the~CSM density and~velocity. Finally, for~the cases where $\tau_{delay}$ =~2~and~8~Myr, we~simulate this delay time by turning  off the~fast wind in~the~inner boundary of our grid and~ letting the~circumstellar  structure evolve for the~relevant time intervals. The~results of our simulations are~illustrated in~Figure~\ref{fig:CSM_PN}.

The moment the~fast wind ceases (Figure~\ref{fig:CSM_PN}a) the~CSM around the~progenitor system is~characterized by a~typical bipolar PN. The~inner cavity of the~PN that was shaped by the~fast wind, is~surrounded by the~dense halo of the~PN---i.e.,~the remnant of the~AGB wind---and~in the~boundary of those two regions a~shell of shock wind is~formed. Such a~structure hosts very similar properties with the~CSM of SNe~Ia that reveal time variable Na ID emission lines (e.g., PTF 2011kx~\cite{Dilday12}) which have been found to~be surrounded by circumstellar shell(s) placed at a~distance of $10^{16} - 10^{17}$~cm from the~explosion center. In~addition, SNRs Ia that reveal profound evidence of interaction with dense CSM such as Kepler's SNR appear most likely to~have occurred in~such a~circumstellar structure (see~Section~\ref{sect:Kepler}).

 For the~case where $\tau_{delay}$ = 2 Myr (Figure~\ref{fig:CSM_PN}b), despite the~fact that~the stellar wind has ceased, due to~the~high radial momentum of the~fast wind, the~circumstellar structure will keep on expanding forming an extended low density cavity of $\sim$7~pc around the~progenitor system. Such a~case offers a~natural explanation on how the~large cavities, observed around several SNRs Ia (e.g., Tycho's SNR, RCW86) can be~formed without requiring the~accretion wind~mechanism.

 Finally, as~time progresses ($\tau_{delay}$ = 8 Myr; Figure~\ref{fig:CSM_PN}c) the~circumstellar structure passes from pressure-driven phase to~the~momentum-driven phase and~inevitably collapses under the~pressure of the~surrounding medium. The~cavity deforms and~the~CSM around the~progenitor system starts~homogenizing. Thus, if~the SN Ia occurs a~couple of tenths of Myr after the~cessation of the~fast wind, no essential circumstellar structures are~expected to~be observed to~its vicinity, in~agreement with a~number of SNe Ia and~their remnants that show no evidence of interaction with~CSM.

%%%%%%%%%%%%%%%%%%%%%%%%%%%%%%%%%%%%%%%%%%
\section{Kepler's SN: A~SNR Resulting by the~Interaction of the~SN Ejecta with the~Surrounding~PN}\label{sect:Kepler}

\begin{figure}[H]
\centering
\includegraphics[width=14cm]{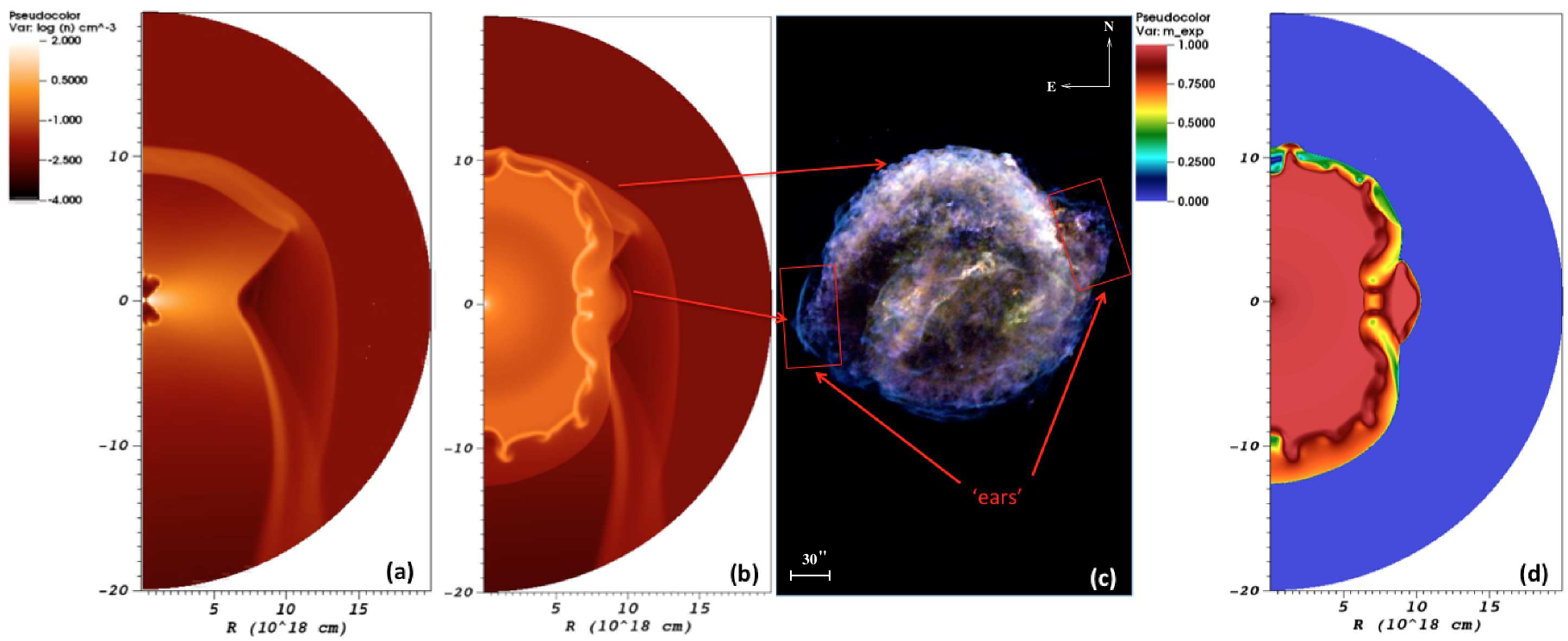}
\caption{(\textbf{a}): the~density distribution of the~CSM that surrounds Kepler's SNR at the~moment of the~explosion. (\textbf{b}) the~density contours of the~SNR after 420 yr of evolution. (\textbf{c}) The~X-ray image of Kepler's SNR \citep{Reynolds07}. (\textbf{d}) the~expansion parameter of the~SNR at t = 420~yr. }
\label{fig:Kepler_sim}
\end{figure}

The Galactic SN of Kepler (SN 1604) is~one of the~best-studied young SNRs (Figure \ref{fig:Kepler_sim}c). There~is~a~consensus that the~remnant resulted by a~SN Ia explosion and~currently it~is~interacting with a~dense circumstellar shell in~its northern region~\cite{Reynolds07}. The~mass contained in~the~shell has been estimated to~be $ > 1\rm M_{\odot}$ and~its chemical composition reveals elevated nitrogen abundances \mbox{($[N/N_{\odot}] > 2$)} \cite{Blair91}. The~interaction between the~SN ejecta and~the~CSM has substantially affected the~dynamics of the~SNR, where in~the~northern region its expansion parameter is~$m= V\times(R/t)= 0.35$, much lower than the~overall expansion of the~remnant which is~m = 0.6~\cite{Vink08}. The~morphology of the~SNR is~rather spherical, revealing optically bright nebulosity in~the~northern portion of the~remnant and~in some central regions due to~the~interaction with the~CSM. The~spherical symmetry of Kepler's SNR breaks in~the~equator of the~remnant, where it~reveals two synchrotron X-ray bright protrusions that give the~impression of two `ears' in~its overall morphology (Figure~\ref{fig:Kepler_sim}c). Finally, based on the~proper motion of the~nitrogen-rich knots~\cite{Bandiera91} and~the~${\rm H}_{\alpha}$ narrow component of the~remnant~\cite{Sollerman03}, it~has been found that Kepler's SNR is~moving with a~high spatial velocity of $u_*$$\approx$$250~ \rm km~s^{-1}$ towards the~north.

Chiotellis~et~al. (2012) \cite{Chiotellis12} modeled Kepler's SNR within the~SN Ia framework. The~authors reproduced the~morphology and~kinematics of the~historical remnant suggesting that the~observed CSM has been shaped by the~slow, nitrogen-rich wind of an AGB donor star, member of the~progenitor~system. The~existence of an AGB shaped CSM around the~remnant was confirmed by the~infrared observations of the~SNR which revealed strong silicate dust features~\cite{Williams12}. Subsequently,~\cite{Patnaude12} modeled the~X-ray spectrum of Kepler's SNR and~found that it~can also be~reproduced considering an evolution of the~SNR within an AGB wind bubble as long as a~small cavity of radius $r$$\sim$0.1~pc is~added in~the~inner region of the~CSM around the~explosion center. Finally,~\cite{Burkey13} in~order to~explain the~observed shocked CSM at the~central regions of the~SNR suggested that the~AGB wind bubble had a~bipolar shape with high mass concentration at the~equatorial plane. Nevertheless,~\cite{Kerzendorf14}
searched at the~center of Kepler's SNR and~found no surviving AGB donor star, confuting the~conclusions of the~previously suggested~models.

The demands imposed by Kepler's SNR observations and~theoretical modeling seem to~be perfectly aligned with a~bipolar PN origin of the~CSM that surrounds the~remnant. In~particular, considering that Kepler's SN occurred inside a~bipolar PN formed by its parent system, we~can naturally and~self-consistently explain: (a) the~current interaction of the~SNR with an AGB wind bubble (i.e.,~the~PN's halo), something that explains the~observed chemical composition of the~CSM and~its properties in~the~IR band, (b) the~small cavity of $r$$\sim$0.1 pc that surrounds the~explosion center needed to~reproduce the~X-ray spectra, which corresponds to~the~inner region of the~PN where the~fast wind dominates, (c) the~density enhancement of the~CSM at the~equatorial region of the~SNR and~finally, (d)~such a~model does not require the~existence of a~survived AGB donor star at the~center of the~SNR. The~only condition demanded by this model is~that the~SN Ia occurred shortly after the~formation of the~PN ($\tau_{delay} \le$ 0.1 Myr ). Such a~demand favors for the~core-degenerate scenario which suggests that the~SN Ia is~triggered by the~merge of a~WD with the~newly born AGB core of the~companion star~\cite{Kashi11}.

Encouraged by the~remarkable similarities between bipolar PNe properties and~these of CSM around Kepler's SNR, we~performed hydrodynamic simulations modeling the~historical remnant within the~framework of the~suggested model. We~first simulated the~formation of the~bipolar PN following the~procedure described in~Section \ref{sec:PN_CSM}. In~order to~include the~observed systemic motion of Kepler's SNR in~our modeling, we~performed our simulations in~the~rest frame of the~progenitor system and~we set the~ISM of density $\rho_{i}$ as an inflow entering the~grid antiparallel the~\emph{y}-axis with a~momentum $m= \rho_{i}u_*cos\theta$, where $u_*= 250~ \rm km~s^{-1}$ (i.e.,~Kepler's SNR systemic velocity). The~resulting circumstellar structure consists of a~typical `hourglass' PN, surrounded by a~bipolar halo formed by the~AGB wind (Figure~\ref{fig:Kepler_sim}a). The~bipolarity of the~halo has been deformed by the~systemic motion of the~progenitor system where a~bow shock is~shaped by the~interaction of the~AGB wind with the~ISM flow. Subsequently, we~introduce in~the~center of this circumstellar structure the~SN ejecta with energy $E_{ej}= 1.2 \times 10^{51}$~erg and~mass $M_{ej}= 1.38 \rm M_{\odot}$ and~we let the~SNR evolve and~interact with the~surrounding medium. Around 420 yr after the~explosion (i.e.,~the~current age of Kepler's SNR) the~largest portion of the~remnant is~well within the~circumstellar structure (Figure~\ref{fig:Kepler_sim}b). However, in~its northern region the~blast wave has reached and~collided with the~bow shaped wind shell. In~addition, at~the equatorial plane of the~SNR, the~forward shock has penetrated the~CSM and~has broken out into the~lower density ambient medium. As~a~result, a~protrusion is~formed in~this region, something that explains the~morphological peculiarity of Kepler's of the~two antisymmetric `ears'. Figure~\ref{fig:Kepler_sim}d depicts the~expansion parameter ($m$) of the~remnant. The~portion of the~SNR that remains within the~CSM reveals an expansion parameter of $m= 0.6$, while in~its northern region that interacts with the~AGB bow shell the~remnant has been substantially decelerated with $m= 0.35$. These values are~consistent with the~results from the~X-ray observations of Kepler's SNR. Finally, the~highest expansion parameter corresponds to~the~region of the~`ears' ($m= 0.8$) in~alignment to~the~intense X-ray synchrotron emission that is~observed at the~two antisymmetric lobes of Kepler's~SNR. 

%%%%%%%%%%%%%%%%%%%%%%%%%%%%%%%%%%%%%%%%%%
\section{Summary}
We have presented evidence that the~ambient medium shaped by PNe can account for several observables of the~CSM that have been found to~surround SNe Ia and~their remnants. The~critical parameter that determines the~diverse CSM properties of SNe/SNRs Ia is~the~delay time between the~PN formation and~the~consecutive SN Ia explosion. This parameter can naturally explain the~existence or~absence of circumstellar structures around the~explosion center as well as the~proximity and~the~density of these~structures. 

Subsequently, motivated by the~intriguing similarities between the~properties of a~bipolar PN and~these of the~CSM that surrounds Kepler's SNR, we~performed 2D hydrodynamic simulations modeling the~historical remnant under the~framework of the~SNe Ia - PN interaction model. We~show that such a~scenario reproduces the~observational characteristics of Kepler's SNR, namely its interaction with a~dense nitrogen-rich circumstellar shell at its northern region, the~current kinematics of the~remnant and~the~existence of the~two antisymmetric lobes on its equatorial~plane. 

%%%%%%%%%%%%%%%%%%%%%%%%%%%%%%%%%%%%%%%%%%
\vspace{12pt} 

{\bf Supplementary material:}  We provide two mp4 simulations of the CSM formation by PNe and of the interaction of Kepler's SNR with a bipolar PN at the electronic version of this article.  

%Please provide

%%%%%%%%%%%%%%%%%%%%%%%%%%%%%%%%%%%%%%%%%%

%%%%%%%%%%%%%%%%%%%%%%%%%%%%%%%%%%%%%%%%%%
\vspace{12pt} 

\authorcontributions{The following statements should be used ``Conceptualization, A.C. and P.B.; methodology, A.C.; software, A.C.; validation, A.C., P.B. and Z.T.S.; formal analysis, A.C.; investigation, A.C., P.B. and Z.T.S; writing--original draft preparation, A.C.; writing--review and editing, A.C., P.B. and Z.T.S.; visualization, A.C.; supervision, P.B.; project administration, P.B.; funding acquisition, A.C., P.B. and Z.T.S. }

%Please provide

%%%%%%%%%%%%%%%%%%%%%%%%%%%%%%%%%%%%%%%%%%
\funding{This research is co-financed by Greece and the European Union (European Social Fund-ESF) through the Operational Programme ``Human Resources Development, Education and Lifelong Learning 2014-2020'' in the context of the project ``On the interaction of Type Ia Supernovae with Planetary Nebulae'' (MIS 5049922).}

%%%%%%%%%%%%%%%%%%%%%%%%%%%%%%%%%%%%%%%%%%
\acknowledgments{AC and~PB gratefully acknowledge the~Lorentz Center for the~support and~the~SOC \& LOC for organizing a~wonderful~workshop.}

%%%%%%%%%%%%%%%%%%%%%%%%%%%%%%%%%%%%%%%%%%

% Please provide either the~correct journal abbreviation (e.g., according to~the~βList of Title Word Abbreviationsβ http://www.issn.org/services/online-services/access-to-the-ltwa/) or~the~full name of the~journal.
% Citations and~References in~Supplementary files are~permitted provided that they also appear in~the~reference list here. 

%=====================================
% References, variant A: external bibliography
%=====================================
%\externalbibliography{yes}
\reftitle{References}

% the~following MDPI journals use author-date citation: Arts, Econometrics, Economies, Genealogy, Humanities, IJFS, JRFM, Laws, Religions, Risks, Social Sciences. For those journals, please follow the~formatting guidelines on http://www.mdpi.com/authors/references
% to~cite two works by the~same author: \citeauthor{ref-journal-1a} (\citeyear{ref-journal-1a}, \citeyear{ref-journal-1b}). This produces: Whittaker (1967, 1975)
% to~cite two works by the~same author with specific pages: \citeauthor{ref-journal-3a} (\citeyear{ref-journal-3a}, p. 328; \citeyear{ref-journal-3b}, p.475). This produces: Wong (1999, p. 328; 2000, p. 475)

%% for journal Sci
%\reviewreports{\\
%Reviewer 1 comments and~authorsβ response\\
%Reviewer 2 comments and~authorsβ response\\
%Reviewer 3 comments and~authorsβ response
%}

%%%%%%%%%%%%%%%%%%%%%%%%%%%%%%%%%%%%%%%%%%
\end{document}